\documentclass[a4paper,12pt]{article}
\usepackage[margin=2cm]{geometry}
\usepackage{comment}
\usepackage{graphicx}
\usepackage{fancyhdr}
\usepackage[colorlinks = true,
            linkcolor = black,
            urlcolor  = blue,
            citecolor = blue,
            anchorcolor = blue]{hyperref}
\usepackage[compat=1.1.0]{tikz-feynhand}
\usepackage{wrapfig}
\usepackage{a4wide}
\usepackage{float}
\usepackage{lscape}
\usepackage{amsmath}
\usepackage{amsfonts}
\usepackage{amssymb}
\usepackage{tikz}
\usepackage[latin1]{inputenc}
\usetikzlibrary{calc,positioning}
\usepackage{mathtools}
\usepackage{lscape}
\usepackage{subcaption}
\usepackage[most]{tcolorbox}
\usepackage{mathtools}
\numberwithin{equation}{section}
\usepackage{authblk}
\usepackage{cite}
\usepackage{slashed}
\usepackage{tensor}
\usepackage[toc,page]{appendix}
\usepackage{dirtytalk}
\usepackage{titlesec}
\usepackage{orcidlink}

\medskip
\newmuskip\pFqmuskip

\newcommand*\pFq[6][8]{%
  \begingroup 
  \pFqmuskip=#1mu\relax
  \mathcode`\,=\string"8000
  \begingroup\lccode`\~=`\,
  \lowercase{\endgroup\let~}\pFqcomma
  {}_{#2}F_{#3}{\left[\genfrac..{0pt}{}{#4}{#5};#6\right]}%
  \endgroup
}
\newcommand{\pFqcomma}{\mskip\pFqmuskip}

\newcommand{\be}{\begin{equation}}
\newcommand{\bea}{\begin{eqnarray}}
\newcommand{\eea}{\end{eqnarray}}
\newcommand{\ba}{\begin{array}}
\newcommand{\ea}{\end{array}}
\newcommand{\ee}{\end{equation}}

\begin{document}
		\begin{center}
{\Large {\bf Logarithmic Corrections for Near-extremal Kerr-Newman Black Holes}}
\vspace*{5mm}
\vspace{10mm}

{\bf \large Ashes Modak\orcidlink{0009-0005-4392-7337}\footnote{\fontsize{8pt}{10pt}\selectfont\ \href{mailto:22dr0061@iitism.ac.in}{22dr0061@iitism.ac.in}}, Aditya Singh\orcidlink{0000-0002-2719-5608}\footnote{\fontsize{8pt}{10pt}\selectfont\ \href{mailto:24pr0148@iitism.ac.in}{24pr0148@iitism.ac.in}} and Binata Panda\footnote{{\fontsize{8pt}{10pt}\selectfont\ \href{mailto:binata@iitism.ac.in}{binata@iitism.ac.in}}\\\;\\ \textbf{\;\;\;\;\;\;\;\;\;\; All Authors Contributed Equally}}}

  \vspace{0.5cm}
  {\it Department of Physics, Indian Institute of Technology (Indian School of Mines), Dhanbad, Jharkhand-826004, India}
  \end{center}

\vspace*{10mm}

\vspace{10mm}

\begin{abstract}
    In this paper, we have computed the logarithmic corrections of entropy for the near-extremal Kerr-Newman black holes in $\mathcal{N}=2$ supergravity theory applying the Euclidean path integral approach in the near-horizon geometry. In the near-horizon extremal Kerr geometry, analogous to the $AdS_{2} \times S^2 $ structure, there exists a set of normalizable zero modes associated with reparametrizations of boundary time. The one-loop approximation to the Euclidean near-horizon extremal Kerr partition function exhibits an infrared divergence due to the path integral over these zero modes. Carrying out the leading finite temperature correction in the near-horizon extremal Kerr scaling limit, we control this divergence. Considering the near-extremal near-horizon geometry as a perturbation around the extremal near-horizon geometry, we determine these corrections implementing a modified heat kernel approach which involves both the extremal and near-extremal corrections and is novel in the literature for the charged rotating black holes in supergravity theory. This result should be reproduced by any microscopic theory that explains the entropy of the black hole.
\end{abstract}
\clearpage
\tableofcontents

\section{Introduction}
Perhaps the closest thing one might observe, in the ongoing pursuit for a quantum theory of gravity could be the black hole thermodynamics\cite{Carlip:2005xy}. The derivation of Hawking temperature and the Bekenstein-Hawking entropy can be considered sufficiently strong that any alleged quantum theory that failed to reproduce these findings would be met with extreme skepticism, even if Hawking radiation has not yet been directly observed\cite{Hawking:1974rv}. We must first tackle a broad but rather sensitive problem before moving on: how can one pose a query concerning a black hole in a quantum theory of gravity? Since the ``usual" response--fixing a black hole background and then asking about quantum fields, gravitational perturbations and the like--seems straightforward, this topic is rarely posed. However, this is not possible in a full quantum theory of gravity because there is no fixed backgrounds and we are unable to merely impose a black hole metric due to uncertainty relations. In order to respond, we must first determine how one requires the presence of a defining property (not the entire classical metric!). Replicating the known thermodynamic features of black holes is a crucial test for any quantum theory of gravity. In the past, a statistical mechanical explanation of the Bekenstein-Hawking entropy looked far-fetched. Despite counting radically different states, the entropy is the same across numerous approaches to quantum gravity\cite{Bombelli:1986rw}. According to this ``universality", the quantum density of states might be governed by some fundamental aspect of the classical theory\cite{Wald:1999vt}. However, we see that in quantum gravitational theories, the entropy of the black hole generally have a universal structure which proves to be of utmost importance.

Akin to the regular thermodynamic systems, black hole entropy ought to be microscopic and specified in context of degeneracy of states in quantum theory\cite{Strominger:1996sh,Balasubramanian:2022gmo,Cvetic:1998sp,Ashtekar:1997yu}. In the framework of string theory, the microscopic counting is well-established for some types of charged black holes\cite{Mandal:2010cj}, notably extremal black holes. A charged black hole consists of an inner (Cauchy) horizon and an event horizon at some temperature which is nonzero, and can be commonly referred to as non-extremal black hole\cite{Mohaupt:2000mj}. In general, non-extremal black holes emit thermal radiations leading to a lower energy or ground state thus rendering the black hole to be extremal\cite{Ferrara:1995ih}. Further, an extremal black hole can be characterized by a zero temperature where the event horizon coincides with the inner horizon. Extremal black holes exhibit some unique properties, such as the vanishing Hawking temperature and connections to supersymmetric solutions in string theory\cite{Strominger:1996kf}. The presence of an infinitely long AdS$_2$ throat near the horizon resulting in an augmented isometry have been well investigated\cite{Hawking:1998jf,Bardeen:1999px}. Also, beyond the semi-classical limit, the logarithmic corrections for these black holes were computed which have been in agreement with microscopic results\cite{Banerjee:2023quv,Banerjee:2023gll,Carlip:2000nv,Strominger:1997eq}. Extremal black holes may boost the understanding of the microstructures of black holes\cite{Singh:2023hit,Singh:2020tkf,Singh:2023ufh}. For various cases\cite{Iliesiu:2022onk,H:2023qko,Gupta:2014hxa,Sen:2012dw,David:2021qaa,Sen_2012}, the logarithmic terms in entropy of the black holes are already computed yet the microscopic descriptions are not easily accessible. The black hole entropy is proportional to the horizon area in the semiclassical regime. This semi-universal entropy form gets the leading order quantum correction that rely on the logarithm of the size of horizon which eventually is dependent on the infrared data of the theory.

Quantum corrections to the black hole entropy formula have been the subject of impressive research efforts in recent years where the quantum-corrected entropy has been expressed as\cite{Karan_2019},
\begin{equation}
    S_B = \frac{A_{H}}{4G_N} + c\ln\frac{A_H}{G_N} + \text{constant} + \mathcal{O}(A_{H}^{-1}).
\end{equation}
The term $c\ln (A_H/G_N)$ represents logarithmic corrections proportional to the logarithm of the black hole's horizon area $A_H$. The terms $\mathcal{O}(A_{H}^{-1})$ denote power-law corrections to the leading-order entropy. However, a detailed microscopic theory of black hole entropy is still lacking, necessitating the ultraviolet (UV) completion of gravity theories\footnote{While total quantum corrections depend on the ultraviolet completion details, logarithmic corrections are determined solely by low-energy modes}. In the black hole entropy, the area and logarithmic terms should, however, be accurately reproduced by any plausible UV-complete theory of gravity. String theory offers a microscopic explanation for the entropy of certain supersymmetric black holes in the extremal limit. For these black holes in various string theory frameworks, the logarithmic corrections to entropy derived from both gravitational and microscopic approaches show remarkable agreement. Recent advancements extend this understanding by reproducing extremal black hole results through the finite-temperature geometry limit\cite{Gupta:2014hxa,Sen:2012dw,David:2021qaa,Sen_2012}. These developments highlight the consistency between microscopic string theory descriptions and macroscopic gravitational computations, reinforcing the robustness of holographic principles and the AdS/CFT correspondence in explaining black hole thermodynamics\cite{Bekenstein:1973ur}. This duality bridges quantum gravity and thermodynamics, offering deeper insights into the nature of black holes\cite{Horowitz:1996qd}.\\

Following recent studies\cite{Bardeen:1999px,Banerjee:2023quv,Banerjee:2023gll,Sen_2012,Karan_2019,Bekenstein:1973ur,Bhattacharyya_2012}, near-extremal black holes possess distinct properties from extremal black holes whose dynamics are effectively governed by a $1D$ Schwarzian theory, with one-loop corrections proportional to the logarithm of temperature differing from the usual logarithmic corrections. In an approach akin to the standard logarithmic corrections, recent studies in\cite{Banerjee:2023gll}, trace back the evolution of these terms\footnote{The effect of small fluctuations are pointed out by these terms, which arise from one-loop quantization on the near-horizon, near-extremal background reflecting a slight shift from the extremal geometry} that accounts to a 4D Euclidean gravity computation. Applying first-order perturbation theory, \cite{Banerjee:2023gll,Kapec:2023ruw,Castro:2021csm} computes eigenvalues of the kinetic operator for small fluctuations near the classical background, uncovering the origin of temperature dependent logarithmic corrections. The extremal limit of black holes potentially violates the Third Law of Thermodynamics and close to extremality, the semiclassical thermodynamic description in the near-extremal limit breaks down\cite{Israel:1986gqz,Liberati:2000zz}. Stated otherwise, even a single Hawking quantum emitted can significantly change the temperature of the near-extremal black holes defying the thermodynamic equilibrium assumption\cite{Preskill:1991tb,Heydeman:2020hhw}. To ascertain the actual low temperature behavior in this regime, quantum correction must be incorporated. A low temperature expansion applying the gravitational path integral was investigated in \cite{Sen:2007qy,Iliesiu:2022kny,Maxfield:2020ale,Gross:1989vs}, to tackle the concern of quantum effects near extremality. Although, the full higher dimensional path integral might not be precisely computed, one can define a set of metric modes\footnote{These modes and their fluctuation determinants are highly sensitive to the type of symmetry preserved by the exact extremal solution such as supersymmetry and other global symmetries} strongly coupled at low temperatures and might not be treated by the semiclassical approximation solely. The aforementioned analysis\cite{Iliesiu:2022kny,Maxfield:2020ale}, demonstrate that at sufficiently low temperatures, the semiclassical thermodynamics is indeed invalid. In the absence of supergravity the low temperature fluctuations remove the ground state degeneracy rendering the degree of state to zero as the energy decreases above extremality. One may perform the gravitational path integral in certain cases, which yields coarse-grained information for the spectrum of microstate of the black hole, provided the corrections are subleading in the extremal entropy and the temperature. A black hole (near extremality) cannot be interpreted as an ordinary quantum system without taking into consideration the low temperature corrections considered in this paper, which are a distinctive fingerprint of quantum gravity and a significant physical effect.

Earlier, logarithmic corrections to the entropy of extremal black holes have been explored following different perspectives dividing broadly into a microscopic and macroscopic class\cite{Banerjee:2010qc}. One used specific microscopic description of the theory for computing the logarithmic corrections in the microscopic class while in the macroscopic case, fluctuating quantum fields were considered in the background for the computation of logarithmic corrections\cite{Sen_20122}. In our approach, we directly compute the entropy considering the quantum fluctuations of different fields in the geometry of black holes. We compute the heat kernel of various fields in the near horizon geometry which is well defined after adopting the infrared subtraction method considered in \cite{Banerjee:2023gll,Sen_2012,Sen_20122}. In order to correctly compute the logarithmic corrections to the entropy of extremal black holes with supersymmetry, \cite{Banerjee:2008ky} used the quantum entropy function formalism where the presence of AdS$_2$ factor in the near horizon geometry of extremal black holes are considered following the AdS$_2$/CFT$_1$ correspondence rules. Notably, this approach is applicable to extremal black holes and is not limited to supersymmetric black holes serving as a versatile tool for investigating quantum corrections to entropy of the black holes. It requires determining the one-loop effective action of massless fields in the near horizon geometry to precisely compute the logarithmic corrections of the black hole. Although, numerous methods have been used for analyzing the effective action, the heat kernel method seems to be most suited and reliable making it particularly advantageous for preserving symmetry. Additionally, it serves as a powerful tool to regularize the UV divergences inherent in the one-loop effective action, ensuring precise and consistent results. As a consequence, there are a number of alternative ways for computing the heat kernel, such as quasinormal modes\cite{Berti:2009kk,Horowitz:1999jd,Denef:2009kn}, group theoretic\cite{Giombi:2008vd,Diaz:2007an}, eigenfunction expansion and specific perturbative expansion methods. Nevertheless, nearly all of the above methods rely on the background geometry or involves some complex computations. 

{\bf Aim and motivations:} The aim is to determine the logarithmic corrections for near-extremal Kerr-Newman black hole in $\mathcal{N}=2$ supergravity theory incorporating the zero modes, non-zero modes and slightly non-zero modes arising from the near-extremal limit of the black hole. The results for the bosonic sector are reevaluated from \cite{Karan_2019,Bhattacharyya_2012} and reviewed but the fermionic sector computations are novel. The computation of the bosonic sector agrees with the eigenfunction approach results of \cite{Sen_2012,Karan_2019,Bhattacharyya_2012} and the hybrid approach results of \cite{Charles:2015eha,Larsen:2014bqa} as well. Further, the logarithmic correction to the near-extremal black hole entropy are computed in \cite{Banerjee:2023quv,Banerjee:2023gll} following the quantum entropy function formalism. We compute this in the near-extremal limit for Kerr-Newman black hole in $\mathcal{N}=2$ supergravity theory stated in eq.(\ref{Log Z}) which involves both the extremal and near-extremal corrections and is novel in the literature. While logarithmic corrections have been extensively studied for extremal black holes, there is a relative paucity of results for near-extremal cases, especially within $\mathcal{N}=2$ supergravity. Supergravity offers the required framework to take into consideration the contributions of different fields and their supersymmetric companions in the context of black hole entropy, ensuring that logarithmic correction computations are in agreement with the underlying symmetries of the theory. Additional complexity in performing entropy corrections is introduced by the rotation\footnote{Rotational degrees of freedom have an impact on the black hole background perturbation spectrum and near-horizon geometry} of Kerr-Newman black holes. In order to fully describe black hole thermodynamics, it is necessary to comprehend how rotation affects logarithmic corrections. Unlike the classical Bekenstein-Hawking entropy alone, these corrections provide information that is sensitive to the spectrum of massless fields and their interactions in the low-energy effective theory.

The organization of this paper is as follows. In section-(\ref{Heat kernel}), we review the heat kernel formalism applying one-loop effective action. In section-(\ref{Kerr Newman}), we determine the logarithmic corrections for extremal Kerr-Newman black hole incorporating zero and non-zero modes only. Section-(\ref{Near extremal KN}), involves our novel computations of the logarithmic corrections for the near-extremal Kerr-Newman black holes in $\mathcal{N}=2$ supergravity theory which describe the main results and is not present in the literature yet. Finally, we summarize with discussion in section-(\ref{Discussion}).

\section{Heat kernel approach: one-loop effective action}\label{Heat kernel}

For the $4D$ Euclidean path integral representation, the partition function $\mathcal{Z}$ is written as,

\begin{equation}\label{Partition function}
    \mathcal{Z}=\int \mathcal{D}[g,{\epsilon}] \exp(-\mathcal{S}[g,{\epsilon}]),
\end{equation}
where, $\mathcal{D}[g,\epsilon]$ is indicating that we have carry out the integration over all possible matter fields $\epsilon$ and metric $g$, and $\mathcal{S}[g,\epsilon]$ is the Euclidean action with the Lagrangian density $\mathcal{L}$ of the matter fields $\epsilon$ propagating through a geometry that is described by the metric g over an arbitrary manifold $\mathcal{M}$ expressed below as,
\begin{equation}
    \mathcal{S}[g,\epsilon]=\int_{\mathcal{M}}d^{4}x\sqrt{\det g} \mathcal{L},
\end{equation}
It is anticipated that the leading contribution to the path integral (\ref{Partition function}) occurs precisely from the metric and the matter fields which are near to the background fields $\Bar{g}$ and $\Bar{\epsilon}$. To incorporate quantum correction, one needs to fluctuate the background field $\Bar{g}$ and $\Bar{\epsilon}$ by quantum fluctuations $\Tilde{g}$ and $\Tilde{\epsilon}$ respectively as,

\begin{equation}
    g=\bar{g}+\tilde{\epsilon}, \quad \epsilon =\bar{\epsilon}+\tilde{\epsilon},
\end{equation}

and further on expanding the action in a Taylor series around the classical solution $(\Bar{g},\Bar{\epsilon})$,

\begin{equation}
    \mathcal{S}[g,\epsilon]=\mathcal{S}[\bar{g},\bar{\epsilon}]+\mathcal{S}_{2}[\tilde{g},\tilde{\epsilon}]+\text{higher order terms}.
    \end{equation}
where, $\mathcal{S}[\Bar{g},\Bar{\epsilon}]$ denotes the action for the classical background fields $\Bar{g}$, and $\Bar{\epsilon}$ and $\mathcal{S}_{2}[\Tilde{g},\Tilde{\epsilon}]$ represents the quadratic
order fluctuated action in the fluctuations $\Tilde{g}$ and $\Tilde{\epsilon}$. The quadratic term of the action is written as\cite{Karan_2019},

\begin{equation}
    \mathcal{S}_{2}[\tilde{g},\tilde{\epsilon}]=\int_{\mathcal{M}}d^{4}x\sqrt{\det\bar{g}}\tilde{\epsilon}_{m} \Delta \tilde{\epsilon}_{n}
\end{equation}

where $\Delta$ is a differential operator that controls the quadratic order quantum fluctuation. Further, the one loop effective action can be written as,

\begin{equation}
    W=\frac{\chi}{2}\ln \det\Delta=\frac{\chi}{2}\sum_{i}\ln\lambda_{i}, \label{one loop action}
\end{equation}
Here, $\chi=+1$ for bosons, $\chi=-1$ for fermions and

\begin{equation}
    \Delta f_{i}=\lambda_{i}f_{i} \label{Eiganvalue eq.}
\end{equation}
Here, ${f_{i}}$ represent the eigenfunctions, and ${\lambda_{i}}$ denote the eigenvalues of the operator $\Delta$. From Eq.~\eqref{one loop action}, it is evident that $\det(\Delta)$ exhibits a divergence due to ultraviolet (UV) contributions. To properly regulate this divergence, we employ the well-established `Heat Kernel' approach, a widely used technique for handling divergences in quantum field theory and spectral analysis.

\subsection{Heat kernel formalism}\label{Heat kernel method}
The heat kernel method serves as an effective and systematic approach for regulating the ultraviolet (UV) divergence that arises in the one-loop effective action \eqref{one loop action}. This technique is particularly useful in spectral analysis and quantum field theory, offering a well-defined framework for handling divergences in functional determinants. Furthermore, by eliminating heavy fields, the heat kernel approach plays a crucial role in obtaining effective field theories. One can extract the contributions of heavy fields to the low-energy effective action, including higher-dimensional operators that capture the effects of the heavy fields, by computing the heat kernel for heavy fields and carrying out the proper-time integration. As established in \cite{Sen_2012,Sen_20122}, the proper-time representation of the heat kernel is given by,
\begin{equation}
    K^{mn}(x,y;s)=\sum_{i}e^{-\lambda_{i}s}f_{i}^{m}(x)f_{i}^{n}(y). \label{Bosonic Heat kernal}
\end{equation}
Here, we choose two points $x$ and $y$ on $\mathcal{M}_{i}$ and $m,n$ are the set of indices to represent eigenfunctions at the points $x$ and $y$ respectively. $s$ denotes a heat kernel parameter, known as the proper time. The eigenfunctions $\{f_{i}^{m}\}$ have been normalized as,

\begin{equation}
    \sum_{i}G_{mn}f_{i}^{m}(x)f_{i}^{n}(y)=\delta^{4}(x,y), \quad \int_{\mathcal{M}}d^{4}x\sqrt{\det\bar{g}}G_{mn}f_{i}^{m}(x)f_{j}^{n}(x)=\delta_{ij}.
\end{equation}
In the space of eigenfunctions, we define the metric $G_{mn}$ which is induced by $\Bar{g}$. We can now express the heat kernel $K(x,y;s)$ as,

\begin{equation}
    K(x,y;s)=G_{mn}K^{mn}(x,y;s),
\end{equation}

which satisfies the diffusion equation,

\begin{equation}
    \left(\frac{\partial}{\partial s}+\Delta_{x}\right) K(x,y;s)=0.
\end{equation}

The operator $\Delta_{x}$ indicates that $\Delta$ is acting only on the first argument of $K(x,y;s)$ and there is a boundary condition $K(x,y;0)=\delta(x,y)$. If we substitute $x=y$ and and integrate over $\mathcal{M}$ then we can define the trace of heat kernel $K(x,y;s)$, known as the heat trace $D(s)$ as,

\begin{equation}
    D(s)=\int_{\mathcal{M}}d^{4}x \sqrt{\det \bar{g}} K(x,x;s)=\sum_{i}e^{-\lambda_{i}s}
\end{equation}
The heat trace $D(s)$ encodes information about the spectrum of the kinetic operator $\Delta$ which is significant in quantum field theory and differential geometry. The eigenvalues of $\Delta$, which are essential for interpreting the geometric properties of manifolds and the behavior of quantum fields, can be extracted by examining $D(s)$. This approach is a prerequisite for investigating spectral invariants and computing one-loop quantum corrections. To calculate the logarithmic partition function in terms of heat kernel, identity is used,

\begin{equation}
   \log(A\xi)=-\lim_{\xi \rightarrow 0} \int_{\xi}^{\infty}\frac{ds}{s}e^{-As} \label{2.13}
\end{equation}

where $\xi \rightarrow 0$ is the UV cutoff of the theory, used to regulate the integral.  Using the identity \eqref{2.13} one can compute the bosonic zero modes\footnote{Zero modes are solutions of a system's equations of motion with zero eigenvalue, or zero energy excitations, in quantum field theory. In order to comprehend the physical characteristics of both bosonic and fermionic systems, these modes are essential.} as\cite{Sen_2012},

\begin{eqnarray}\label{bosonic zero modes}
\log \mathcal{Z}_{nz}=-\frac{1}{2}\sum_{i}^{nz}e^{-\lambda_{i}s} =\frac{1}{2}\int_{\xi}^{\infty}\frac{ds}{s}\int d^{4}x\sqrt{\det \bar{g}}K^{nz}(x,x;s)
\end{eqnarray}


The non-zero contributions are denoted by $nz$. The one-loop contribution to the partition function $\mathcal{Z}_{AdS_{2}}$ from non-zero mode is given by the product of the eigenvalues $\lambda_{i}$ of the kinetic operator and on further taking the natural logarithm yields:
\begin{equation}
    \prod_{i}^{nz}\lambda_{i}^{-\frac{1}{2}}=e^{-\frac{1}{2}\ln \lambda_{i}}.
\end{equation}
This form is significant as it simplifies the computation of the one-loop effective action by converting the product over eigenvalues into a sum, facilitating the analysis of quantum corrections in the AdS$_2$ background. As wave functions are orthogonal, we can rewrite eq. \eqref{bosonic zero modes} as,

\begin{equation}
   \log \mathcal{Z}_{nz}=\frac{1}{2}\int_{\xi}^{\infty}\frac{ds}{s}\int d^{4}x\sqrt{\det \bar{g}}\left(K(0;s)-\bar{K}(0)\right) \label{2.15}
\end{equation}

where $K((0;s)=G_{mn}K^{mn}(x,x;s)$ and $\Bar{K}(0)=G_{mn}\Bar{K}^{mn}(x,x;s)$. 

Here, $\Bar{K}^{mn}(x,y)=\sum_{l} g_{l}^{m}(x)g_{l}^{n}(y)$, if $g_{l}^{m}$'s are any special set of $f_{i}^{m}$'s for which $\lambda_{i}$ vanishes. If we consider the fermionic fluctuation\cite{Banerjee:2023gll},

\begin{equation}
    K_f^{mn}(x,y;s)=-\frac{1}{2}\sum_{i}e^{-\lambda_{i}^{f}s}f_{i}^{m}(x)f_{i}^{n}(y)
\end{equation} \label{Fermionic HK}
In fermionic fluctuations, the heat kernel contains an additional $-\frac{1}{2}$ term and the absolute form remain same. The fermionic zero-mode contribution can be written as, 

\begin{eqnarray}\label{2.17}
  \log \mathcal{Z}_{nz}=\frac{1}{2}\int_{\xi}^{\infty}\frac{ds}{s}\int d^{4}x\sqrt{\det \bar{g}}\left(K(0;s)-\bar{K}_{f}(0)\right)
\end{eqnarray}

\section{Extremal Kerr-Newman black holes in \texorpdfstring{$\mathcal{N}$} \texorpdfstring{=2} supergravity theory} \label{Kerr Newman}

The general form of the metric for Kerr-Newman black hole is given as \cite{Bhattacharyya_2012},
\begin{equation}\label{Kerr Newman metric}
\begin{split}
ds^{2}&=-\frac{(r^{2}+a^{2}\cos^{2}\psi)(r^{2}+a^{2} \cos^{2}\psi-2Mr+Q^{2})}{(r^{2}+a^{2}\cos^{2}\psi)(r^{2}+a^{2})+(2Mr-Q^{2})a^{2}\sin^{2}\psi}dt^{2}\\
&+\frac{(r^{2}+a^{2}\cos^{2}\psi)}{(r^{2}+a^{2} \cos^{2}\psi-2Mr+Q^{2})}dr^{2}+(r^{2}+a^{2}\cos^{2}\psi)d\psi^{2}\\
&+\frac{(r^{2}+a^{2}\cos^{2}\psi)(r^{2}+a^{2})+(2Mr-Q^{2})a^{2}\sin^{2}\psi}{(r^{2}+a^{2}\cos^{2}\psi)}\sin^{2}\psi\\
&\times \left(d\phi + \frac{(Q^{2}-2Mr)a}{(r^{2}+a^{2}\cos^{2}\psi)(r^{2}+a^{2})+(2Mr-Q^{2})a^{2}\sin^{2}\psi}dt\right)^{2}\\
\end{split}
\end{equation}

Here taking $G_N=1/16\pi $, the mass $m$, charge $q$ and angular momentum $J$ conjugate to the parameter $M$, $Q$ and $a$ can be written respectively as,

\begin{equation}
    m=16\pi M, \quad q=8\pi Q, \quad J=16 \pi aM 
\end{equation}
The location of the horizon can be given as,
\begin{equation}
    r^{2}-2Mr+a^{2}+Q^{2}=0
\end{equation}
In the near horizon geometry\cite{Bhattacharyya_2012}, the metric \eqref{Kerr Newman metric} can be written as,

\begin{equation}
\begin{split}
    ds^{2}&=\bar{g}_{\mu\nu}dx^{\mu}dx^{\nu}\\
    &=(Q^{2}+a^{2}+a^{2}\cos^{2}\psi)(d\eta^{2}+\sinh^{2}\eta d\theta^{2}+d\psi^{2})\\
    &+\frac{(Q^{2}+2a^{2})^{2}}{(Q^{2}+a^{2}+a^{2}\cos^{2}\psi)}\sin^{2}\psi \left(d\chi-i\frac{2Ma}{Q^{2}+2a^{2}}(\cosh\eta-1)d\theta\right)\label{3.4}
\end{split}
\end{equation}

Utilizing eq.\eqref{3.4}, it is straightforward to evaluate
\begin{equation}
    \sqrt{\det\bar{g}}=G(\psi)\sinh\eta=(Q^{2}+a^{2}+a^{2}\cos\psi^{2})(Q^{2}+2a^{2})\sin\psi \sinh\eta
\end{equation}

\subsection{Bosonic mode}
Utilizing eq.\eqref{2.15}, we can evaluate the non-zero mode contribution

\begin{equation}
   \log \mathcal{Z}_{nz}=\frac{1}{2}\int_{\xi}^{\infty}\frac{ds}{s}\int d^{4}x\sqrt{\det \bar{g}}\left(K(0;s)-\bar{K}(0)\right) \label{3.6}
\end{equation}
In order to express the above integral in terms of $\ln \Lambda$ where $Q,a \sim \Lambda$ and $A_H \sim \Lambda^2$, we have to go to the region $1<<s<<\Lambda^{2}$ in the $s$ integral. Further, on series expansion of $K(0;s)$ in $\Bar{s}=s/ \Lambda^{2}$ region around $\Bar{s}=0$ and considering $K_{0}$ as coefficient of constant mode in this expansion, we can rewrite eq.\eqref{3.6} as follows \cite{Sen_20122},

\begin{eqnarray}\label{3.7}
   \log \mathcal{Z}_{nz}&=&\frac{1}{2}\int d^{4}x\sqrt{\det \bar{g}}\left(K_{0}-\bar{K}(0)\right) \ln \Lambda^{2}\\
   &=&8\pi^{2} \left(Q^4 +\frac{10a^{2}Q^{2}}{3}+\frac{8a^{4}}{3}\right)(\cosh \eta_{0}-1)\left(K_{0}-\bar{K}(0)\right) \ln \Lambda.
\end{eqnarray}

The range of the near-horizon coordinates are $0<\eta<\eta_{0}$ and $0<\theta<2 \pi$. The location of the horizon is at $\eta = 0$. One can note that, in the limit where angular momentum is vanishing i.e., $a=0$, the above expression matches with that of extremal Reissner-Nordstr\"{o}m (RN) black holes given as,
\begin{equation}
     \quad \int d^{4}x\sqrt{\det \bar{g}}=8\pi^{2}Q^{4}(\cosh \eta_{0}-1).
\end{equation}
Also, taking $Q=0$, we get the expression for near-extremal Kerr black holes
\begin{equation}
    \quad \int d^{4}x\sqrt{\det \bar{g}}=\frac{64\pi^{2}Q^{4}}{3}(\cosh \eta_{0}-1).
\end{equation}

In field space, the matrix $\bar{K}^{mn}$ assumes a block diagonal structure, where each block corresponds to the zero modes associated with distinct sets of fields, denoted by $\{A_r\}$. For a given set $\{A_r\}$, the number of zero modes can be expressed as:
\begin{eqnarray}
 \nonumber       N_{zm}^{(r)}&=&\int d^{4}x \sqrt{\det \bar{g}}\bar{K}^{r(0)} \\        &=& 8\pi^{2}\left(Q^4 +\frac{10a^{2}Q^{2}}{3}+\frac{8a^{4}}{3}\right)(\cosh\eta_{0}-1)K^{r}(0)\\
        \bar{K}^{mn}(x,y)&=&\sum_{l \in A_{r}} G_{mn} g_{l}^{m}(x)g_{l}^{n}(y)
\end{eqnarray}

These zero modes are inherently linked to certain asymptotic symmetries, particularly gauge transformations characterized by parameters that remain non-vanishing at infinity. To integrate over the zero modes, one can redefine variables by expressing the coefficients of these modes in terms of the parameters governing the (super)-group of asymptotic symmetries. Specifically, for zero modes associated with the $r$'th block, this transformation introduces a Jacobian factor proportional to $a^{\beta_{r}}$ for each mode. As a result, the overall dependence of $Z_{AdS_{2}}$ on $a$, stemming from the integration over zero modes, is dictated by this factor.
\begin{equation}
    a^{ \sum_{r} \beta_{r} N_{zm}^{r}} = e^{ 8\pi^{2} \left(Q^4 +\frac{10a^{2}Q^{2}}{3}+\frac{8a^{4}}{3}\right) (\cosh \eta_{0} - 1) \ln \Lambda  \sum_{r} \beta_{r} \bar{K}^{r}(0)}
\end{equation}

Thus, the contribution of the zero modes to the logarithm of the partition function is expressed as follows,

\begin{eqnarray}
\nonumber    \ln Z_{zm}&=&\sum_{r}\beta_{r}N_{zm}^{r}\ln \Lambda\\
    &=&8\pi^{2} \left(Q^4 +\frac{10a^{2}Q^{2}}{3}+\frac{8a^{4}}{3}\right) (\cosh \eta_{0}-1)\ln \Lambda \sum_{r} \beta_{r}\bar{K}^{r}(0). \label{3.13}
\end{eqnarray}

 The coefficient of $\cosh \eta_{0}$ can be understood as arising from a shift in the energy $E_{0}$ while the term independent of $\eta_{0}$ is interpreted as a contribution to the black hole entropy. From eq.\eqref{3.7} and eq.\eqref{3.13}, the total logarithmic correction for bosonic sector is computed as,

 \begin{equation}
 \begin{split}
     \ln \mathcal{Z}_{Bosons}&\sim -8\pi^{2}\left(Q^4 +\frac{10a^{2}Q^{2}}{3}+\frac{8a^{4}}{3}\right)\left(K_{0}+\sum_{r} (\beta_{r}-1)\bar{K}^{r}(0)\right) \ln{\Lambda}\\
     & \sim -4\pi^{2}\left(Q^4 +\frac{10a^{2}Q^{2}}{3}+\frac{8a^{4}}{3}\right)\left(K_{0}+\sum_{r} (\beta_{r}-1)\bar{K}^{r}(0)\right) \ln A_{H}\\
     \end{split}
 \end{equation}

 If the angular momentum of the black hole is set to zero ($a = 0$), it reduces to Reissner-Nordstr\"{o}m black hole. In this limit, we obtain the scaling relation $ \Lambda \sim Q $.
 \begin{equation} \label{Bosonic RN Blak Hole}
 \begin{split}
     \ln \mathcal{Z}_{Bosons}& \sim -8\pi^{2}Q^4\left(K_{0}+\sum_{r} (\beta_{r}-1)\bar{K}^{r}(0)\right) \ln{\Lambda}\\
     &\sim -4\pi^{2}Q^4\left(K_{0}+\sum_{r} (\beta_{r}-1)\bar{K}^{r}(0)\right) \ln A_{H}\\
     \end{split}
 \end{equation}
This equation \eqref{Bosonic RN Blak Hole} precisely corresponds to the results obtained for the extremal Reissner-Nordstr\"{o}m black hole in the bosonic sector \cite{Sen_20122}.

\subsection{Fermionic mode}
For the fermionic fluctuation, we can perform an analysis exactly similar to the bosonic fluctuation. Following eq.\eqref{2.17}, we can write the non-zero contribution as,

\begin{eqnarray}\label{3.15}
\nonumber    \log \mathcal{Z}_{nz}&=&\frac{1}{2}\int d^{4}x\sqrt{\det \bar{g}}\left(K_{0}^{f}-\bar{K}^{f}(0)\right) \ln \Lambda^{2}\\
    &=& 8 \pi^{2} \left(Q^4 +\frac{10a^{2}Q^{2}}{3}+\frac{8a^{4}}{3}\right) (\cosh \eta_{0}-1)\left(K_{0}^{f}-\bar{K}^{f}(0)\right) \ln \Lambda 
\end{eqnarray} 

On evaluating the number of zero modes in terms of the  trace of heat kernel, we get

\begin{eqnarray} \label{Number of Fermionic Zero modes}
 \nonumber   N_{zm}^{(f)}&=&-2\int d^{4}x \sqrt{\det \bar{g}}\bar{K}^{f}(0)\\
    &=&-16 \pi^{2}\left(Q^4 +\frac{10a^{2}Q^{2}}{3}+\frac{8a^{4}}{3}\right)(\cosh\eta_{0}-1)\bar{K}^{f}(0)\\ 
    \bar{K}^{mn}(x,y)&=&-\frac{1}{2}\sum_{l \in A_{r}} G_{mn}^{f} g_{l}^{(f)m}(x)g_{l}^{(f)n}(y)
\end{eqnarray}

Let us further assume that integrating over each fermion zero mode contributes a factor proportional to $a^{-\beta_{f}/2}$, where $\beta_{f}$ is a constant. Consequently, the total contribution dependent on $a$ from the integration over all fermion zero modes can be expressed as,
\begin{equation}
    e^{8\pi^{2}\left(Q^4 +\frac{10a^{2}Q^{2}}{3}+\frac{8a^{4}}{3}\right) (\cosh \eta_{0}-1)\ln \Lambda\sum_{f} \beta_{f}\bar{K}^{f}(0)}.
\end{equation}
Hence, the zero-mode contribution to the logarithm of the partition function can be represented as follows,

\begin{equation}
    \ln Z_{zm}=8\pi^{2} \left(Q^4 +\frac{10a^{2}Q^{2}}{3}+\frac{8a^{4}}{3}\right) (\cosh \eta_{0}-1)\ln \Lambda \sum_{f}\beta_{f}\bar{K}^{f}(0). \label{3.18}
\end{equation} 

As mentioned earlier, the coefficient of $\cosh \eta_0$ is interpreted as resulting from a shift in the energy $E_0$, whereas the term that does not depend on $\eta_0$ is associated with a contribution to the black hole entropy. From eq.\eqref{3.15} and eq.\eqref{3.18}, the total logarithmic correction for the fermionic sector is computed as,

\begin{equation}
\begin{split}
\ln\mathcal{Z}_{Fermions}&\sim -8\pi^{2}\left(Q^4 +\frac{10a^{2}Q^{2}}{3}+\frac{8a^{4}}{3}\right)\left(K_{0}^{f}+\sum_{f} (\beta_{f}-1)\bar{K}^{f}(0)\right) \ln \Lambda\\
&\sim -4\pi^{2}\left(Q^4 +\frac{10a^{2}Q^{2}}{3}+\frac{8a^{4}}{3}\right)\left(K_{0}^{f}+\sum_{f} (\beta_{f}-1)\bar{K}^{f}(0)\right) \ln A_{H}\\
\end{split}
 \end{equation}

 If the black hole's angular momentum vanishes ($a = 0$), it transitions into a Reissner-Nordstr\"{o}m black hole. In this regime, the scaling relation $ \Lambda \sim Q $ holds.

 \begin{equation} \label{Fermionic RN Blak Hole}
 \begin{split}
     \ln \mathcal{Z}_{Fermions}& \sim -8\pi^{2}Q^4\left(K_{0}+\sum_{f} (\beta_{r}-1)\bar{K}^{f}(0)\right) \ln{\Lambda}\\
     &\sim -4\pi^{2}Q^4\left(K_{0}+\sum_{f} (\beta_{f}-1)\bar{K}^{f}(0)\right) \ln A_{H}\\
     \end{split}
 \end{equation}
This equation \eqref{Fermionic RN Blak Hole} perfectly agrees with the results for the extremal Reissner-Nordstr\"{o}m black hole in the fermionic sector.

\section{Near-extremal Kerr-Newman black holes in \texorpdfstring{$\mathcal{N}$} \texorpdfstring{=2} supergravity theory}\label{Near extremal KN}
In the $T\rightarrow 0$ limit, corresponding to near-extremality, we analyze the black hole by determining the eigenvalues of the kinetic operator and computing the one-loop corrected partition function. The near-horizon region is treated as a small temperature perturbation of the extremal geometry. For extremal black holes, the infinitely extended near-horizon throat confines computations to this region. Thus, we focus on the near-horizon physics, where the near-extremal geometry is described by a linear-order temperature deviation of \( AdS_2 \times \mathcal{K} \).

\begin{figure}[ht]
    \begin{center}
        \includegraphics[scale=1.0]{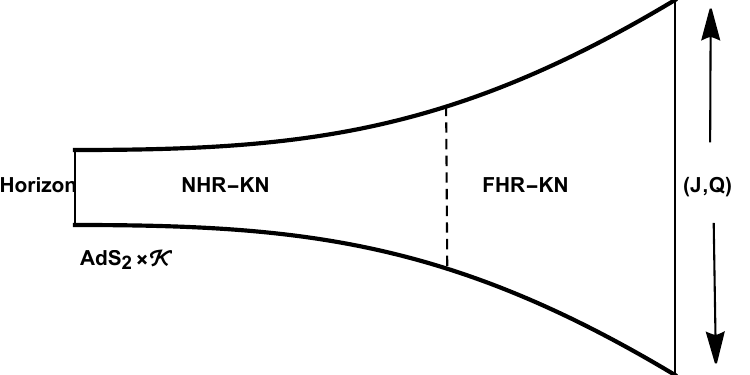}
    \end{center}
    \caption{A schematic representation of the near-horizon region (NHR) and far-horizon region (FHR) for a near-extremal Kerr-Newman (KN) black hole in $\mathcal{N}=2$ supergravity theory. The NHR-KN region is characterized by the $AdS_2 \times \mathcal{K}$ geometry.}
    \label{fig:KN_horizon}
\end{figure}

To analyze this system, we quantize it on the near-horizon background using first-order perturbation theory, as outlined in \cite{Banerjee:2023quv}. However, instead of relying on explicit eigenvalue calculations only, we adopt a heat kernel method that bypasses such computations \cite{Banerjee:2023gll}. The fundamental idea behind the first-order perturbation theory approach is to reinterpret the quantization procedure around the near-extremal background by analyzing the spectrum-specifically, the heat kernel of a modified operator-within the $AdS_{2} \times \mathcal{K}$ geometry. Given that the near-extremal background represents a small linear deviation in temperature from the extremal state, the quadratic action can be naturally reformulated within this framework:
\begin{equation}
     \mathcal{S}_{2}[\tilde{g},\tilde{\epsilon}]=\int_{\mathcal{M}}d^{4}x\sqrt{\det\bar{g}}\tilde{\epsilon}_{m}(\Delta+T\Delta^{(c)})\tilde{\epsilon}_{n}.
\end{equation}
We are interested in determining the heat kernel for the changed operator as we have combined all of the temperature-dependent adjustments into the operator $\Delta^{(c)}$. Considering $\Bar{\Delta}=\Delta+T\Delta^{c}$ and utilizing eq.\eqref{Eiganvalue eq.}, we can write $\Bar{\Delta}\Bar{f}_{i}=\Bar{\lambda}_{i}\Bar{f}_{i}$
where
\begin{equation}
    \bar{\lambda}_{i}=\lambda_{i}+T\lambda_{i}^{(c)}, \quad\bar{f}_{i}=f_{i}+Tf_{i}^{(c)}.
\end{equation}

Using the first-order perturbations theory, we obtain

\begin{equation}
    \lambda_{i}^{(c)}=\int d^{4}x\sqrt{\det \bar{g}} f_{i}^{m}(x)\Delta_{mn}^{(c)}f_{i}^{n(x)}
\end{equation}

\begin{equation}
    f_{i}^{(c)}(x)=\sum_{j\neq i}\frac{1}{\lambda_{i}-\lambda_{j}}\left(\int d^{4}x^{\prime}\sqrt{\det \bar{g}} f_{j}^{n}(x^{\prime})\Delta_{np}^{(c)}f_{i}^{p}(x^{\prime})\right)f_{j}^{m}
\end{equation}

Using these eigenvalues and eigenfunctions, the heat kernel is again defined in \eqref{Bosonic Heat kernal} for bosonic fluctuations and \eqref{Fermionic HK} for fermionic fluctuations. Our objective is to extract the logarithmic corrections in $\Lambda$ and $T$, as these introduce distinct scales to the solution. Following the similar approach\cite{Banerjee:2023gll}, one can compute the total logarithmic correction for the  near-extremal Kerr-Newman black holes as,

\begin{equation} \label{Exp of total ln Z}
    \ln \mathcal{Z}= \ln \mathcal{Z}_{nz}+\ln \mathcal{Z}_{snz}+\ln \mathcal{Z}_{zm}
\end{equation}
where, $ \ln \mathcal{Z}_{nz} $ and $ \ln \mathcal{Z}_{zm} $ represent the contributions arising respectively from the proper non-zero modes and the zero modes of the extremal black hole. Consequently, they only contribute to the $\ln \Lambda$ term. Due to the near-extremal consideration, there arises another kind of non-zero mode called slightly non-zero modes denoted as $\ln \mathcal{Z}_{snz}$ and the contribution is coming from these modes. The slightly non-zero modes were originally zero modes on the extremal background but became non-zero due to the introduction of temperature.\\

Zero modes in an extremal black hole are intimately connected to the spontaneous breaking of asymptotic symmetries near the $AdS_2$ boundary. When a small temperature is introduced, these zero modes acquire a mass gap (i.e., they get lifted), and the underlying symmetries are modified. The near-horizon geometry of an extremal black hole exhibits an $SL(2,{R})$ symmetry, which governs the structure of the $AdS_2$ region. While this symmetry is globally preserved in the extremal limit, fluctuations around the background extend it to an infinite-dimensional Virasoro algebra, corresponding to large diffeomorphisms at the $AdS_2$ boundary. Additionally, the presence of gauge symmetries-such as $U(1)$ for charge and $SO(3)$ for rotational symmetry-further enhances the asymptotic structure, leading to large gauge transformations. However, these extended symmetries are spontaneously broken, resulting in an infinite number of associated zero modes. Among these zero modes, three classes are particularly significant: tensor modes, $l = 0$ vector modes, and $l=1$ vector modes, each representing distinct perturbations of the black hole background. As demonstrated in \cite{Banerjee:2023gll}, when the black hole is perturbed away from extremality, only the tensor modes and $l=1$ vector modes-which correspond to metric fluctuations-persist as slightly non-zero modes. This lifting of zero modes characterizes the transition from an extremal to a near-extremal black hole, encoding fundamental changes in the underlying symmetry structure.

\subsection{Bosonic slightly non-zero mode}
The contribution of the slightly non-zero modes will be treated separately, as these modes contribute to both $ \ln \Lambda $ and $ \ln T $ terms. The partition function for slightly non-zero modes is given as,
\begin{equation}
   \ln \mathcal{Z}_{snz}=\frac{1}{2}\int_{\xi}^{\infty}\frac{1}{2}\left(\overline{\sum}_i e^{-T\lambda_{i}^{c}s}\right)
\end{equation}

We use the notation $\overline{\sum}_{i}$ to denote the summation over the zero modes of the extremal solution, which are elevated to slightly non-zero modes. The final expression of the  integration (following \cite{Banerjee:2023gll}) :

\begin{equation}
    \ln \mathcal{Z}_{snz}=-\frac{1}{2}N_{snz} \ln \left(\frac{T \xi}{\Lambda} \right)
\end{equation}

It follows that the coefficient of this correction is determined by the number of zero modes in the extremal solution that transition to non-zero modes upon introducing a small temperature.

\begin{equation}
    \ln \mathcal{Z}_{snz}=8\pi^{2}\left(Q^4 +\frac{10a^{2}Q^{2}}{3}+\frac{8a^{4}}{3}\right)(\cosh\eta_{0}-1)\tilde{K}^{r}(0)
\end{equation}

The tilde denotes a sum over slightly non-zero modes. As the term that does not depend on $\eta_0$ is associated with a contribution to the black hole entropy, the final expression for $\ln N_{snz}$ is given by,
\begin{equation}
     \ln \mathcal{Z}_{snz}^{Bosons}\sim4\pi^{2}\left(Q^4 +\frac{10a^{2}Q^{2}}{3}+\frac{8a^{4}}{3}\right)\tilde{K}^{r}(0) \ln \left(\frac{T\xi}{\Lambda} \right)
\end{equation}

If the black hole is non-rotating ($a = 0$), it reduces to a near-extremal Reissner-Nordstr\"{o}m black hole, where the scaling relation $ \Lambda \sim Q $ holds. In this regime, we can further derive the following.

\begin{equation} \label{Bosonic snz for RN}
     \ln \mathcal{Z}_{snz}^{Bosons}\sim4\pi^{2}Q^4\tilde{K}^{r}(0) \ln \left(\frac{T\xi}{\Lambda} \right)
\end{equation}
This equation \eqref{Bosonic snz for RN} accurately corresponds to the results obtained for the bosonic slightly non-zero mode sector of a near-extremal Reissner-Nordstr\"{o}m black hole\cite{Banerjee:2023gll}.

\subsection{Fermionic slightly non-zero mode}
The computation of fermionic slightly nonzero modes differs from that of bosonic modes. In this case, the relevant kinetic operator is linear in derivatives rather than quadratic\footnote{The eigenvalues of the fermionic kinetic operator generally take the form  
$\lambda_{i}^{f} = \alpha + \gamma T$.  
For slightly nonzero modes, we set $\alpha = 0$. However, since the $\mathcal{O}(T)$ term is proportional to the $\mathcal{O}(1)$ term, it follows that $\gamma = 0$. Consequently, no $\mathcal{O}(T)$ contribution should be present in the spectrum. }. Since the eigenvalues of the squared kinetic operator scale as $\mathcal{O}(T^{2})$, obtaining the corresponding eigenvalues for the slightly nonzero modes requires taking their square root. The contribution of these modes to the partition function is given by,
\begin{equation}
    \ln \mathcal{Z}_{snz}^{f}=\frac{1}{2}\overline{\sum}_{i} \ln \lambda_{i}^{f}
\end{equation}

From the equation \eqref{2.13} we can write,

\begin{equation}
    \ln \mathcal{Z}_{snz}^{f}=-\frac{1}{2}\int_{\xi^{\prime}}^{\infty}\frac{ds}{s}\left(\overline{\sum}_{i}e^{-T\lambda_{i}^{c}s}\right)
\end{equation}

The auxiliary variable used in this context has different dimensions compared to the bosonic case. Here, $\xi^{\prime}$ represents a small UV cutoff of the theory, while $\lambda_{i}^{c}$ characterizes the modification in the eigenvalues of the fermionic kinetic operator \footnote{referring specifically to the operator itself and not its square} associated with the slightly nonzero modes.

To obtain the logarithmic correction, we perform the rescaling $s \rightarrow \hat{s} = T s$.

\begin{eqnarray}
\nonumber     \ln \mathcal{Z}_{snz}^{f}&=&-\frac{1}{2} \int_{T\xi^{\prime}}^{\infty}\frac{d \hat{s}}{\hat{s}}\left(\overline{\sum}_{i}e^{-\lambda_{i}^{c}\hat{s}}\right)\\
     &=&\frac{1}{2}N_{snz}^{f}\ln (T\xi^{\prime})
\end{eqnarray}

Here, $N_f^{{snz}}$ represents the number of fermionic slightly nonzero modes, which is defined by eq.\eqref{Number of Fermionic Zero modes} through a summation over the relevant modes. It is straightforward to observe that, if we were to use the eigenvalues of the squared operator, the lower limit of the $s$-integral would be given by $T^2 \xi$, where $\xi$ denotes the UV cutoff. By making this comparison, we find that $\xi' \sim \sqrt{\xi}$. Therefore, no logarithmic dependence on $\Lambda$ is introduced by this sector.

\begin{equation}
\ln \mathcal{Z}_{snz}^{f}=-8\pi^{2} \left(Q^4 +\frac{10a^{2}Q^{2}}{3}+\frac{8a^{4}}{3}\right)(\cosh \eta_{0}-1)\ln (T\xi^{\prime})
\end{equation}

As the term that does not depend on $\eta_0$ is associated with a contribution to the black hole entropy, the final expression for $\ln \mathcal{Z}_{snz}$ is given by,

\begin{equation}
\ln \mathcal{Z}_{snz}^{Fermions}\sim 8\pi^{2} \left(Q^4 +\frac{10a^{2}Q^{2}}{3}+\frac{8a^{4}}{3}\right)\ln (T\xi^{\prime})
\end{equation}

If the black hole has no rotation $a = 0$, it transitions into a near-extremal Reissner-Nordstr\"{o}m black hole. In this limit, we can further obtain,

\begin{equation} \label{Fermionic snz for RN}
\ln \mathcal{Z}_{snz}^{Fermions}\sim 8\pi^{2} Q^4\ln (T\xi^{\prime})
\end{equation}

This equation \eqref{Fermionic snz for RN} exactly matches the results for the fermionic slightly non-zero mode sector of the near-extremal Reissner-Nordstr\"{o}m black hole. We will now outline the full logarithmic contributions (coming from \eqref{Exp of total ln Z}) in terms of $\Lambda$ and temperature. The sole difference between the extremal and near-extremal cases arises from the sector of slightly non-zero modes.
\begin{equation}\label{Log Z}
\begin{split}
     \ln \mathcal{Z}^{KN}& \sim \left[\mathcal{C}_{r}^{KN}+8\pi^{2}\left(Q^4 +\frac{10a^{2}Q^{2}}{3}+\frac{8a^{4}}{3}\right)\sum_{r \in snz}\left(\beta_{r}-\frac{1}{2}\right) K^{r}(0)  \right]\ln \Lambda\\
     &+\left[\mathcal{C}_{f}^{KN}+8\pi^{2}\left(Q^4 +\frac{10a^{2}Q^{2}}{3}+\frac{8a^{4}}{3}\right)\sum_{f \in snz}\beta_{f} \bar{K}^{f}(0)\right] \ln \Lambda\\
     &+8\pi^{2}\left(Q^4 +\frac{10a^{2}Q^{2}}{3}+\frac{8a^{4}}{3}\right)\sum_{r \in snz} \frac{1}{2}\bar{K}^{r}(0) \ln T\\
     &+8\pi^{2}\left(Q^4 +\frac{10a^{2}Q^{2}}{3}+\frac{8a^{4}}{3}\right)\sum_{f \in snz}\bar{K}^{f}(0) \ln T
    \end{split}
\end{equation}

Here $\mathcal{C}_{r}^{KN}$ and $\mathcal{C}_{f}^{KN}$ are the coefficients of $\ln \Lambda$, for the bosonic and fermionic sections of the extremal black hole respectively and are expressed as,
\begin{equation}
\mathcal{C}_{r}^{KN}=-8\pi^{2}\left(Q^4 +\frac{10a^{2}Q^{2}}{3}+\frac{8a^{4}}{3}\right)\left(K_{0}+\sum_{r} (\beta_{r}-1)\bar{K}^{r}(0)\right)
\end{equation}

\begin{equation}
\mathcal{C}_{f}^{KN}=-8\pi^{2}\left(Q^4 +\frac{10a^{2}Q^{2}}{3}+\frac{8a^{4}}{3}\right)\left(K_{0}^{f}+\sum_{f} (\beta_{f}-1)\bar{K}^{f}(0)\right) 
\end{equation}

To obtain the complete logarithmic correction for a near-extremal Reissner-Nordstr\"{o}m black hole, one should set $a = 0$ (and $\Lambda \sim Q)$. In this limit, the expression for the near-extremal Kerr-Newman black hole with $\mathcal{N}=2$ supergravity theory reduces to the near-extremal Reissner-Nordstr\"{o}m black holes which is given as,

\begin{equation}\label{Log Z RN}
\begin{split}
     \ln \mathcal{Z}^{RN}& \sim \left[\mathcal{C}_{r}^{RN}+8\pi^{2}Q^4\sum_{r \in snz}\left(\beta_{r}-\frac{1}{2}\right) K^{r}(0)  \right]\ln \Lambda\\
     &+\left[\mathcal{C}_{f}^{RN}+8\pi^{2}Q^4\sum_{f \in snz}\beta_{f} \bar{K}^{f}(0)\right] \ln \Lambda\\
     &+8\pi^{2}Q^4\sum_{r \in snz} \frac{1}{2}\bar{K}^{r}(0) \ln T +8\pi^{2}Q^4\sum_{f \in snz}\bar{K}^{f}(0) \ln T
    \end{split}
\end{equation}
where $\mathcal{C}_{r}^{RN}$ and $\mathcal{C}_{f}^{RN}$ can be expressed as,
\begin{equation} \label{CrRN}
\mathcal{C}_{r}^{RN}=-8\pi^{2}Q^4(K_{0}+\sum_{r} \big(\beta_{r}-1)\bar{K}^{r}(0)\big)
\end{equation}

\begin{equation} \label{CfRn}
\mathcal{C}_{f}^{RN}=-8\pi^{2}Q^4\big(K_{0}^{f}+\sum_{f} (\beta_{f}-1)\bar{K}^{f}(0)\big) 
\end{equation}
Eq.\eqref{Log Z} represents our final result for the logarithmic corrections to near-extremal black holes where we have written the novel expression for the logarithmic correction to the entropy of a near-extremal Kerr-Newman black hole in $\mathcal{N}=2$ supergravity theory. This result provides a deeper understanding of the quantum corrections to black hole entropy and highlights the significance of near-extremal geometries in studying such effects. Our computation is based on the precise evaluation of the one-loop determinants arising from the fluctuations around the near-horizon background. The consistency of this result reinforces the validity of our approach and offers valuable insights into the role of supersymmetry in black hole thermodynamics. Moreover, this study lays the groundwork for future investigations into logarithmic corrections in more general black hole solutions within supergravity and string theory frameworks.

\section{Discussion}\label{Discussion}
We computed the final expression of $\ln \mathcal{Z}$ for near-extremal Kerr-Newman black hole given in eq.\eqref{Log Z}. One can note that due to the near-extremal correction, only the metric zero modes has been lifted (see section-\ref{Near extremal KN}). The quantity termed as the \textit{regularized zero modes} can be expressed as,
\begin{equation}
    N_r = -3 - \mathcal{K},
\end{equation}
where $\mathcal{K}$ represents the number of rotational isometries\cite{Bhattacharyya_2012}. For Kerr-Newman black hole $\mathcal{K} = 1$ and thus $N_r = -4$. Following \cite{Charles:2015eha} we can write,
\begin{eqnarray} \label{Relation of K and C_{zm}}
\nonumber    \frac{1}{2}C_{zm}&=&-4\pi^{2}\left(Q^4 +\frac{10a^{2}Q^{2}}{3}+\frac{8a^{4}}{3}\right)\sum_{r}(\beta_{r}-1)\bar{K}^{r}(0)\\
    &-&4\pi^{2}\left(Q^4 +\frac{10a^{2}Q^{2}}{3}+\frac{8a^{4}}{3}\right)\sum_{f}(\beta_{f}-1)\bar{K}^{f}(0)
\end{eqnarray}

Here, $\beta_{r}=2$ and $\beta_{f}=3$ for a four dimensional black hole\cite{Sen_2012} and $C_{zm}$ is defined as, 
\begin{equation}
    C_{zm}= N_r+2N_{SUSY}
\end{equation}
where $N_{SUSY}$ is the number of preserved supercharges (4 for BPS black holes and 0 otherwise). Utilizing this, it is straightforward to compute,
\begin{equation}
    \bar{K}^{r}(0)= \frac{1}{\pi^{2}\left(Q^4 +\frac{10a^{2}Q^{2}}{3}+\frac{8a^{4}}{3}\right)}
\end{equation}
Also following \cite{Karan_2019}, we see that for extremal Kerr-Newman black holes\footnote{See eq.(6.14) in arXiv:1905.13058 [hep-th].}, $C_{zm}=-4$ and thus we obtain,
\begin{equation}
    \bar{K}^{f}(0)=-\frac{1}{4\pi^{2}\left(Q^4 +\frac{10a^{2}Q^{2}}{3}+\frac{8a^{4}}{3}\right)}
\end{equation}
Further, substituting this in eq.\eqref{Log Z} we can write,

\begin{equation} \label{Final Result KN}
\begin{split}
    \ln \mathcal{Z}^{KN}&=[\mathcal{C}_{r}^{KN}+\mathcal{C}_{f}^{KN}+6]\ln \Lambda+2\ln T\\
    &=[\mathcal{C}_{ext}^{KN}+6]\ln \Lambda +2\ln T
    \end{split}
\end{equation}
where $\mathcal{C}_{ext}^{KN}=\mathcal{C}_{r}^{KN}+\mathcal{C}_{f}^{KN}$. Eq.\eqref{Final Result KN} is our final $\ln \mathcal{Z}$ for near-extremal Kerr-Newman black holes in $\mathcal{N}=2$ supergravity theory. From \cite{Sen_20122}, for near-extremal Reissner-Nordstr\"{o}m black hole\footnote{A detailed computation of $\bar{K}^{r}(0)$ and $\bar{K}^{f}(0)$ is provided in section-4 of \cite{Sen_20122}.}
\begin{eqnarray}\label{kr for RN}
    \bar{K}^{r}(0)=\frac{3}{4\pi^{2} Q^{4}}, \quad \bar{K}^{f}(0) = -\frac{1}{2\pi^{2}Q^{4}}
\end{eqnarray}
Utilizing eq.\eqref{Log Z}, eq.\eqref{kr for RN} and further taking the limit $a=0$, one can directly arrive at the expression for logarithmic corrections of near-extremal Reissner-Nordstr\"{o}m black hole in $\mathcal{N}=2$ supergravity theory\footnote{According to \cite{Kaul:2000kf}, the entropy has a logarithmic correction of $-3/2 \ln A_H=-3\ln Q$. However, the area variable $A_H$ counts the number of states per unit interval. Consequently, converting it in terms of $\Lambda$ the result changes.} given as,
\begin{eqnarray}
   \ln \mathcal{Z}^{RN}=[C_{ext}^{RN}-3]\ln Q-\ln T
\end{eqnarray}

Where $\mathcal{C}_{ext}^{RN}=\mathcal{C}_{r}^{RN}+\mathcal{C}_{f}^{RN}$.This result precisely matches the complete logarithmic correction obtained for the near-extremal Reissner-Nordstr\"{o}m black hole in $\mathcal{N}=2$ supergravity (see eq.(4.10) in \cite{Banerjee:2023gll}). Such an agreement serves as a strong validation of our methodology and the accuracy of our computation for the near-extremal Kerr-Newman black hole in pure $\mathcal{N}=2$ supergravity. This consistency reinforces the robustness of our approach in evaluating quantum corrections to black hole entropy and highlights the universality of logarithmic corrections in near-extremal black hole solutions within the supergravity framework. Furthermore, it provides deeper insights into the underlying structure of quantum gravity corrections and their interplay with supersymmetry in extremal black hole configurations.\\

In order to compute the density of states in the regime where the near-extremal partition function is valid, one can apply the inverse Laplace transformation as,
\begin{equation}
    \rho(E) = e^{S_{0}}\int d\beta e^{\beta E+\frac{S_{1}}{\beta}}z(\beta)
\end{equation}
where the saddle point contribution\footnote{One can calculate the saddle point contribution by applying the traditional Gibbons-Hawking-York prescription, as discussed in \cite{Sen:2012dw}.} to the partition function is included and the extremal part is denoted as $S_0$ while $S_1$ corresponds to the near-extremal correction part. The computation is valid in the regime where the bound for temperature is $(\Lambda T)) \ll 1$ and the energy above extremality satisfies,
\begin{eqnarray}
   y\left(\frac{\xi}{\Lambda^2}\right) \ll (\Lambda E) \ll 1
\end{eqnarray}
where, $y$ represents a power of the dimensionless combination $\xi/\Lambda^2$. Since the precise power-law dependence varies depending on the field fluctuations, the largest of these bounds is considered the lower limit. In order to ensure the validity of the analysis, the temperature must continue to be significantly higher than this threshold. The contribution of the one-loop partition function is denoted as $Z(\beta)$ arising from the logarithmic correction and $E$ is the energy just above extremality. Utilizing the Gibbons-Hawking-York prescription, the saddle point contribution can be determined. It is straightforward to compute the partition function utilizing eq.\eqref{Final Result KN} as,
\begin{equation}
    z(\beta)\sim \Lambda^{(C_{ext}^{KN}+6)}\beta^{-2}.
\end{equation}
This is the precise result for the logarithmic correction to the entropy of near-extremal Kerr-Newman black holes which shed light on quantum corrections to black hole entropy. While the supersymmetric index is invariant across parameter changes, entropy measures the degeneracy of states in supersymmetric black hole analysis, allowing comparisons between macroscopic and microscopic approaches.  As discussed in \cite{Dabholkar:2010rm}, the quantum entropy function computes this index, enabling direct comparison with microscopic calculations.\\

For extremal Kerr-Newman black holes in $\mathcal{N}=2$ or $4$ supersymmetric theories in four dimensions, the logarithmic corrections are predicted from the macroscopic point of view but the microscopic results are yet not available. We have computed the logarithmic corrections for the near-extremal Kerr-Newman black holes in $\mathcal{N}=2$ supergravity from the microscopic side. Our analysis ensures consistency between macroscopic entropy and microscopic state counts, providing a robust framework for analyzing supersymmetric black holes. Since the near-extremal regime is inconsistent with traditional methods for non-extremal black holes, our framework offers a key tool for examining these corrections in near-extremal black holes. Any microscopic theory explaining extremal Kerr-Newman black hole entropy is subject to strong constraints following our result stated in eq.(\ref{Final Result KN}), which requires that it  must account for both the leading Bekenstein-Hawking term and the sub-leading logarithmic correction. The search for this underlying CFT could be guided by eq.(\ref{Final Result KN}), especially if the entropy of the extremal Kerr-Newman black hole has an underlying two dimensional conformal field theory, as proposed in\cite{Guica:2008mu}.\\

Our computation closely resembles that of an extremal\footnote{The term ``extremal black holes" corresponds to the extremal-limit of non-extremal black holes, as is earlier highlighted.} black hole but differs from non-extremal cases, even with nonzero temperature. Unlike non-extremal computations, which use the full geometry, near-extremal calculations rely on near-horizon geometry. We reformulate the quantization problem by finding eigenvalues of a modified operator on the extremal background via first-order perturbation theory. In contrast to non-extremal black holes, near-extremal black holes exhibit non-uniform parameter scaling. Our approach analyzes scale-dependent heat kernel contributions from slightly non-zero modes (former zero modes of extremal black holes that become weakly non-degenerate) and isolates logarithmic contributions from inverse temperature and charges. These modes arise from asymptotic symmetry breaking near the AdS$_2$ boundary, while the decoupled asymptotic region introduces no additional zero modes. Metric zero modes are uplifted by near-extremality, while large gauge symmetries associated with charges are not affected. First-order perturbation theory computes reliable logarithmic contributions.  As discussed in section-(\ref{Near extremal KN}), modes that are slightly non-zero at first order only introduce some temperature corrections at higher orders. At higher orders, zero modes at first order could appear to lift, but symmetries stop further logarithmic contributions. Since charges remain fixed, large gauge symmetries are preserved, ensuring accurate logarithmic corrections in the near-extremal regime. Thus, in the near-extremal regime, our results accurately represent the logarithmic corrections which remain consistent within a specific temperature range, constrained by an upper bound related to the charges and a lower bound determined by the UV cutoff theory. This lower cutoff reflects a limitation of current approach.\\

On the other hand, this approach can be applied to calculate logarithmic entropy corrections in near-extremal solutions for any gravity theory. In the heat kernel formalism, the lower limit of the Schwinger parameter integration is modified by a nonzero temperature, preventing it from being strictly set to zero. The coefficient of the logarithmic correction is dependent on the counting of slightly non-zero modes and may vary subject to the theory considered. One cannot employ a zero-temperature limit directly to the final result as such regularization by their very nature enforce a nonzero temperature condition. Consequently, the logarithmic corrections derived from the extremal near-horizon geometry remain valid across any gravity theory. Extremal black holes universally feature an AdS$_2$ component, where large diffeomorphism symmetries are lifted at finite temperatures. This AdS$_2$ structure persists in rotating black holes, implying that the argument remains valid.\\

{\bf Remarks:} The thermodynamics of various black hole solutions are significantly impacted by quantum corrections. The presence of non-renormalization properties in the extremal supersymmetric black holes make them an ideal framework for testing the consistency between macroscopic entropy predictions and the microscopic entropy computation. The general analysis in \cite{Banerjee:2010qc}, describes that an extremal black hole determines the entropy within microcanonical ensemble, where the charges corresponding to all gauge fields in AdS$_2$ are fixed. In addition to the gauge fields obtained from the metric, arising from the rotational isometries of the black hole solution, the gauge fields on AdS$_2$ incorporate all the Maxwell fields of the theory. The computation and physical interpretation of logarithmic corrections to thermodynamics of near-extremal black holes share a number of subtle difficulties. Gauge and geometric ambiguities arise when `fixing' the near-horizon disconnected geometry to the asymptotically flat region of spacetime, is a key limitation. Connecting the two regions is not simple, as it requires matching the whole black hole solution in asymptotically flat space with the symmetries, gauge choices and physical characteristics of the near-horizon geometry. These ambiguities make it more difficult to derive physical insights and to develop a theoretical framework.

\section*{Acknowledgments}
A.M. is thankful to Anowar Shaikh for initial discussion of the work. A.S. would like to thank CSIR-HRDG for financial support received as RA working under Project No. 03WS(003)/2023-24/EMR-II/ASPIRE. B.P. thanks CSIR-HRDG, Govt of India, for financial support received through Grant No. 03WS(003)/2023-24/EMR-II/ASPIRE.

\bibliographystyle{JHEP}
\bibliography{Referencetemplate}

\end{document}